\documentclass[doublecol,linenumbers]{epl2}
\usepackage{epsfig}
\usepackage{dcolumn}
\usepackage{amsmath}
\begin{document}
\title{\bf\ Radiation-induced re-emission  in
a 2D electron system.}

\author{J. I\~narrea}
\institute{Escuela Polit\'ecnica
Superior, Universidad Carlos III, Leganes, Madrid, 28911, Spain}
\pacs{nn.mm.xx}{First pacs description}
\pacs{nn.mm.xx}{Second pacs description}
\pacs{nn.mm.xx}{Third pacs description}

\date{\today}
\abstract{
Recent experiments on re-emission of radiation and  magnetotransport in
photoexcited two-dimensional electron systems are theoretically analyzed.
These experiments have concluded that there exists a strong correlation
between the re-emitted radiation and the radiation-driven current through the sample.
We study these remarkable experimental results using the radiation-driven electron orbit model.
According to it, two-dimensional electrons under a magnetic field move back and forth  driven by
radiation. Therefore, they behave as oscillating dipoles re-emitting part of the radiation
previously absorbed. Then, this theory would explain in a simple way the experimental results.}
\maketitle
\section{Introduction}
Two-dimensional electron systems (2DES) implemented in ultra-high mobility GaAs/AlGaAs
heterostructures are called to be part of the key building blocks in the current and future nanoelectronics.
Their interaction with radiation and the electron transport through them are of
special interest from both theoretical and experimental standpoints.
For instance, those systems are potentially useful for electromagnetic wave
characterization in the microwave (MW) and terahertz bands with a
technological interest that can be considered beyond question. Accordingly, remarkable effects such
as microwave-induced resistance oscillations (MIRO) and zero-resistance
states (ZRS) have drawn much attention from the condensed matter physics community.
These effects
were  discovered  when a high-mobility 2DES in a low and perpendicular magnetic field ($B$)
was irradiated with MW \cite{mani1,zudov1}.
Different
theories have been proposed to explain these  effects
\cite{ina2,ina20,girvin,dietel,lei,rivera,vavilov} but the physical
origin  still remains unclear.
In the same way, an important  experimental effort has  been  made too
\cite{mani2,mani3,willett,mani4,smet,yuan,vk}.

Very recent experimental results\cite{mani5,mani6} on MIRO compare the magnetoresistance ($R_{xx}$)
response of the microwave-excited 2DES with the concurrent re-emission
from the same system that is detected by a nearby carbon resistor sensor\cite{mani5,mani6}.
They report on a strong correlation between the results coming from
both types of measurements. They also study the dependence on MW frequency and
power suggesting a cause and effect relationship between the microwaves
on the 2DES and what is observed in the current through the carbon resistor.
Another important result they obtain is that the re-emission
signal in the carbon sensor remains unchanged even
when the current through the 2DES sample is switched off. The surprising
outcome suggests a radiation-induced change in the electronic properties
of the 2DES even in the absence of applied current.
According to these experiments\cite{mani5,mani6}, the carbon sensor detects both the incident or
primary and the re-emitted radiation. The incident radiation is constant,
thus it produces a magnetic field independent change in the carbon resistor.
On the other hand, the re-emitted  radiation depends on the magnetic field due to
the radiation-induced resistance oscillations and cyclotron resonance, both
in the 2DES. Therefore for experimentalists, it is easy
to tell the difference between incident and re-emitted signals\cite{mani5,mani6}.
\begin{figure}
\centering\epsfxsize=3.2in \epsfysize=4.0in
\epsffile{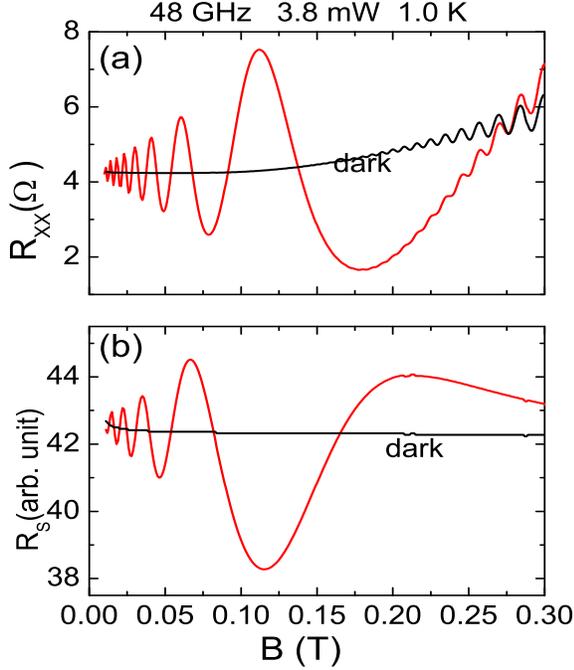}
\caption{(Color online). a) Calculated longitudinal magnetoresistance, $R_{xx}$, vs magnetic field with (red
curve) and without (black curve) 48 GHz, MW irradiation.
b) Calculated magnetoresistance in the carbon resistor, $R_{S}$, versus magnetic field
with (red
curve) and without (black curve) re-emitted radiation from
the 2D GaAs/AlGaAs heterostructure.}
\end{figure}

In this letter, we
theoretically study  the physical origin of these experimental results and their dependence on MW frequency,
 power and dc-current. We focus on the re-emitted radiation and study the concurrence between both
transport data from the GaAs/AlGaAs heterostructure and from the carbon resistor sensor.
We also predict about the re-emission process from the 2DES  and its effect on
the carbon resistor signal when the 2DES undergoes a regime of ZRS. We are based on a previous theoretical model
developed by the authors to deal with
MIRO and ZRS \cite{ina2,ina20,ridley,ina3,ina4,ina5,inainplane}: {\it the radiation (or MW)
-driven electron orbits model}.
According to this theory, two-dimensional electrons under a static magnetic field, i.e., two
dimensional electron orbits, oscillate being  spatially driven by
radiation. Therefore, they behave as oscillating dipoles re-emitting part of the radiation
previously absorbed. Then, this theory would explain in a simple way the experimental results\cite{mani5,mani6}.

\section{Theoretical model}
The above mentioned theory\cite{ina2,ina20,ridley} was proposed to explain
the $R_{xx}$ of an irradiated 2DES at low $B$. We obtained the
exact solution of the corresponding electronic wave
function:
$\Psi(x,t)\propto\phi_{n}(x-X-x_{cl}(t),t)$
, where $\phi_{n}$ is the solution for the
Schr\"{o}dinger equation of the unforced quantum harmonic
oscillator, $X$ is the center of the orbit for the electron
motion, $x_{cl}(t)$ is the classical solution of a forced harmonic
oscillator:
\begin{eqnarray}
x_{cl}(t)&=&\left[\frac{e
E_{o}}{m^{*}\sqrt{(w_{c}^{2}-w^{2})^{2}+\gamma^{4}}}\right]\cos wt\nonumber\\
&=&A\cos wt
\end{eqnarray}
where $e$ is the electron charge, $\gamma$ is a
phenomenologically-introduced damping factor for the electronic
interaction with acoustic phonons, $w_{c}$ the cyclotron
frequency and $E_{0}$ the MW-radiation electric field.
 Then, the
obtained wave function is the same as the standard harmonic
oscillator where the center is displaced by $x_{cl}(t)$. Thus, the
orbit centers are not fixed, but they oscillate harmonically at
the radiation field frequency $w$.
This $radiation-driven$ behavior affects dramatically the charged
impurity scattering and eventually the conductivity. Then, first we
calculate the impurity scattering rate $W_{n,m}$\cite{ina2,ina20,ridley}
between two $oscillating$ Landau states $\Psi_{n}$, and
$\Psi_{m}$. Next we find the average effective distance advanced by the electron
in every scattering jump: $\Delta X^{MW}\propto A\cos w\tau$,
where $\tau=1/W_{n,m}$ is
the scattering time\cite{ina2,ina20,ridley}. Finally the longitudinal conductivity
$\sigma_{xx}$ is given by:
$\sigma_{xx}\propto \int dE \frac{\Delta X^{MW}}{\tau}$
being $E$
the energy.
To obtain $R_{xx}$ we use
the relation
$R_{xx}=\frac{\sigma_{xx}}{\sigma_{xx}^{2}+\sigma_{xy}^{2}}
\simeq\frac{\sigma_{xx}}{\sigma_{xy}^{2}}$, where
$\sigma_{xy}\simeq\frac{n_{i}e}{B}$ and $\sigma_{xx}\ll\sigma_{xy}$.
Finally it turns out that $R_{xx}\propto A$ and then
the amplitude of MIROs mainly depends on $A$.
\begin{figure}
\centering\epsfxsize=3.5in \epsfysize=4.3in
\epsffile{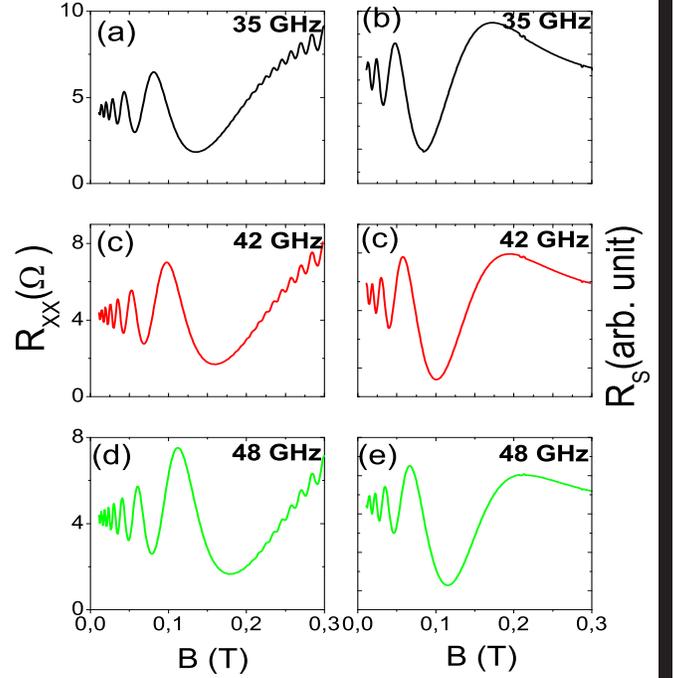}
\caption{(Color online). Left panels show the longitudinal resistance $R_{xx}$
vs the magnetic field $B$. Right panels show the carbon sensor resistance $R_{s}$
vs $B$. Calculated results correspond to different frequencies: 35, 42 and 48 GHz.
For all curves we observe how the oscillatory structures span to higher $B$ as
the MW-frequency increases. In the same way more oscillations are obtained.}
\end{figure}

One of the most important outcomes of the present theory
is that the electron orbits undergo a classical back and forth motion through $x_{cl}$  driven
by the MW. Accordingly, the time-dependent x-coordinate of the guiding center, $X^{MW}(t)$
performs this harmonic motion because $X^{MW}(t)=x_{cl}(t)$.
Therefore, each electron orbit behaves as an oscillating electric dipole
where the dipole moment $\Pi$ changes as $\Pi=eX^{MW}(t)=\Pi_{0}\cos wt$
where $\Pi_{0}=eA$. According to classical electromagnetism, this oscillating dipole is able to re-emit
radiation at the same frequency of MW.
The time dependent electric field $E_{S}$ of this
re-emitted radiation coming out from one of these oscillating dipoles is given by\cite{wag}
\begin{eqnarray}
E_{S}&=&\left[\frac{\Pi_{0}w^{2}}{4\pi\epsilon_{0}\epsilon_{GaAs}z_{0}c_{GaAs}^{2}}\right]\cos wt\nonumber\\
&=&\left[\frac{eA w^{2}}{4\pi\epsilon_{0}\epsilon_{GaAs}z_{0}c_{GaAs}^{2}}\right]\cos wt\nonumber\\
&=&E_{S0}\cos wt
\end{eqnarray}
where $c_{GaAs}$ is the speed of light in GaAs, $\epsilon_{GaAs}$
is the GaAs dielectric constant, $\epsilon_{0}$ is the dielectric constant
in vacuum and $z_{0}$ is distance between the 2DES and the carbon resistor.
\begin{figure}
\centering \epsfxsize=3.2in \epsfysize=4.3in
\epsffile{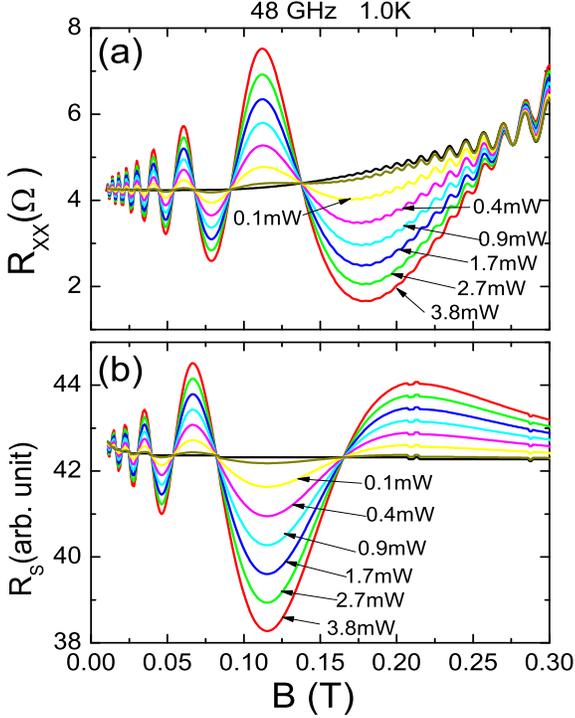}
\caption{  Dependence on the MW-power of the, a) calculated longitudinal magnetoresistance, $R_{xx}$, vs magnetic field,
and b) calculated magnetoresistance in the carbon resistor, $R_{S}$, versus magnetic field.
The MW-power is swept from 3.2 mW to dark. MW-frequency is 48 GHz and temperature T=1.0K.
It is clearly observed that MW illumination affects similarly both types of measurements: the radiation-response
 vanishes  as the MW-power decreases.}
\end{figure}
\begin{figure}
\centering \epsfxsize=3.5in \epsfysize=3.7in
\epsffile{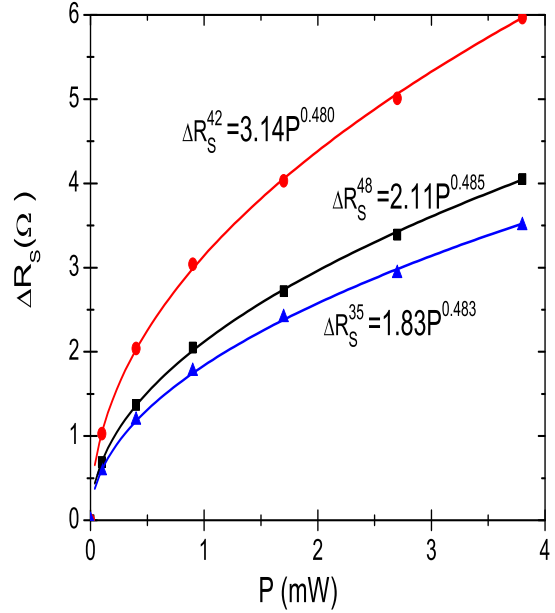}
\caption{Calculated $\Delta R_{S}=R_{S}(E_{ST})-R_{S}(0)$ vs
the MW power $P$, for frequencies of 35, 42, 48 GHz. $R_{S}(E_{ST})$ is the calculated $R_{S}$ for the irradiated carbon sensor
and $R_{S}(0)$ without irradiation. Three corresponding fits show up in the panel suggesting a square root
dependence of $R_{S}$ versus  $P$. }
\end{figure}
The total electric field of the radiation illuminating  the carbon
resistor will be the sum of all individual electric fields coming out
from each oscillating dipole (oscillating electron orbit) in the 2DES.
Since all electron orbits oscillate in phase we can express the
total electric field as,
\begin{equation}
E_{ST}=\sum_{i}(E_{S})_{i}=NE_{S0}\cos wt
\end{equation}
where the
subindex $i$ in the sum denotes the individual oscillating orbits.  $N$ is the total
number of electron orbits contributing to $E_{ST}$, being $N=nS$, where
$n$ is the 2D electron density and $S$ the surface of the sample.
\begin{figure}
\centering \epsfxsize=3.2in \epsfysize=3.5in
\epsffile{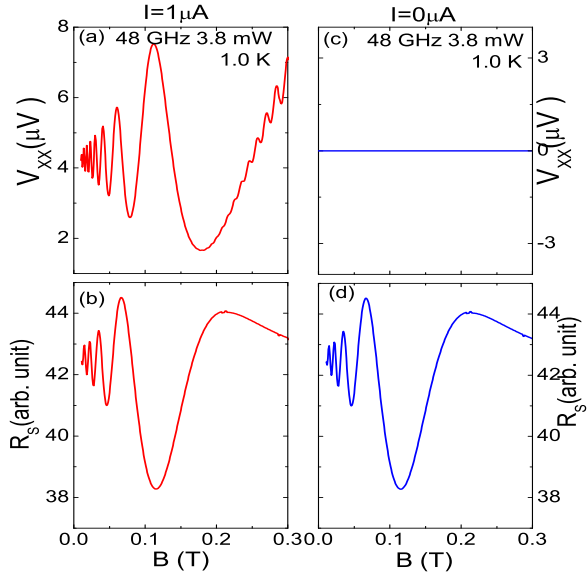}
\caption{Calculated longitudinal voltage $V_{xx}$ and carbon sensor
resistance, $R_{S}$ vs magnetic field for different values of the DC current $I$, under $48$ GHz
MW irradiation. In panels a) and b) the dc-current through the 2DES is
$I=1\mu A$. In  c) and d), $I=0\mu A$. Remarkably,
$R_{S}$ turns out to be immune to the applied current through the sample. }
\end{figure}
This theory obtains, in agreement with experiments, that the electric field amplitude of the re-emitted
radiation, $NE_{S0}$, depends on the magnetic field through $A$ and in turn on $w_{c}=\frac{eB}{m^{*}}$.
 On the other hand, the electric field amplitude, $E_{0}$ of
the primary MW radiation illuminating the sensor, is constant, i.e., independent of
the magnetic field.

\revision{
Equation [3] suggest a remarkable fact as it is that the re-emitted radiation
could be a possible example of {\it superradiance}. Superradiance\cite{dicke} is a cooperative phenomenon
in which the radiation coming from an ensemble of $N$ microscopic emitters in phase with each other
is substantially enhanced. Two main conditions have to be fulfilled in order to obtain superradiance:
first the distance between emitters has to be much smaller than the radiation wavelength and second,
the net electromagnetic field emitted by whole ensemble has to be proportional to $N$. Therefore the
emitted intensity goes as $N^{2}$. Based in our theoretical approach we can state that the
re-emitted radiation in the corresponding experiments\cite{mani5,mani6}, fulfills all the  requirements
to be considered as an example of {\it superradiance}. The distance between Landau states (emitters)
is much smaller than the MW  wave length and acoording to equations [3] the net electromagnetic field
goes as $N^{2}$. This prediction could be easily confirmed in
an experiment with a sample of tunable electron density via a voltage gate. In this way, it could be
studied the dependence of the re-emitted intensity with the electron density, i.e., the number of emitters,
and to check out if intentisty goes as the square of the electron density.}

To calculate
the conductivity and resistance, $R_{S}$, of the carbon resistor under irradiation,
 as a first approach, we extend
the radiation-driven electron orbit model to the resistor.
The electrons inside the resistor contributing to the current
are subjected to the same $B$, turning into quantum oscillators.
Therefore we approximate them as 2D-electrons which are able to
be coupled to radiation and spatially driven by it.
It is not intended in the current letter to present a microscopical model for the current through
the carbon sensor in the presence of a magnetic field. We rather focus  on
how the re-emitted radiation couples with the electrons in the sensor and study  the MW-mediated
strong correlation between carbon sensor and GaAs-2DES.

We expect to obtain, in agreement with experiments, an oscillating
profile for $R_{S}$ versus magnetic field. According to our theory, the electric field of radiation illuminating the
resistor is given by $E_{ST}$, which
implies that $R_{S}\propto E_{ST}$. Then, the oscillation frequency of the electron orbits in the resistor
will be the same as the MW-frequency. On the other
hand, the oscillations amplitude of the
oscillatory structure obtained from the sensor
will mainly depend on $NE_{S0}$, that plays the role of $E_{0}$ for MIROs. This will be
reflected in the radiation response measured in $R_{S}$.
Another
different feature in the case of the resistor is the impurity scattering rate, $W_{S}$.
In our approximation we consider  that the conductivity is worse in
the carbon resistor than in the GaAs-2DES and then the scattering rate
will be bigger and the scattering time smaller:
$W_{S}>W_{n,m}$ and  $\tau_{S}<\tau$, $\tau_{S}$ being the impurity
scattering time in the resistor and related with the scattering rate by
$\tau_{S}=1/W_{S}$. For the numerical calculations we have phenomenologically
assumed that $W_{S}\simeq 2 \times W_{n,m}$ and then
 $ \tau_{S}\simeq 0.5 \times \tau$.
\begin{figure}
\centering \epsfxsize=3.2in \epsfysize=3.3in
\epsffile{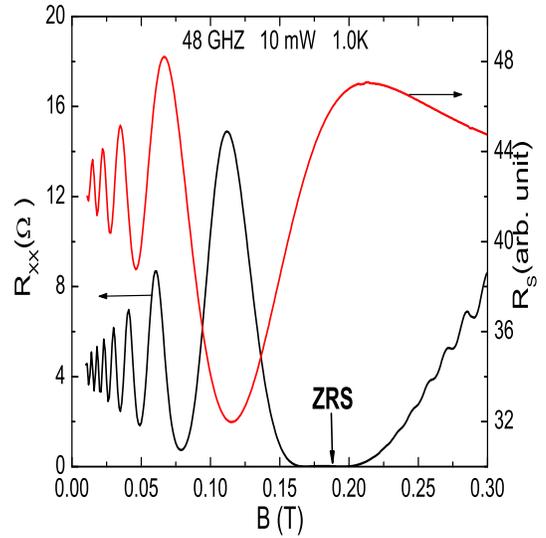}
\caption{ Calculated $R_{xx}$ under
MW-irradiation of high power (left $y$-axis)
and $R_{S}$ (right $y$-axis) versus $B$.
It is observed that the radiation-related features of $R_{S}$ are
not affected by the existence of a ZRS regime in $R_{xx}$
around $B=0.18$ T. }
\end{figure}
\section{Results}
In Fig. 1 we present in the upper panel, calculated diagonal magnetoresistance $R_{xx}$ of
the 2DES and in the lower panel, calculated carbon sensor resistance, $R_{S}$, in both cases
with respect to $B$. The dark case is also presented for both panels, turning out
to be featureless: for the upper panel MIRO are not obtained, as expected, and
for lower panel there is only a flat line.
But when  MW of $48$ GHz is switched on, we observe  MIRO in the upper panel,  and
for the lower one we observe important radiation-related oscillatory features that we can consider
concurrent with the obtained MIRO in the 2DES.
In Fig. 2,
we exhibit $R_{xx}$ and $R_{S}$ responses for different MW-frequencies. Left column presents
$R_{xx}$  vs $B$ for 35, 42 and 48 GHz. We observe that MIROs span to higher $B$ as
the MW-frequency is increased and at the same time more oscillations turn up. In the right
column we present the concurrent carbon sensor response $R_{S}$ vs $B$ for the
same frequencies as in the left column. Similarly as in the experiments and the left column, we
observe that as MW-frequency increases more oscillations show up and
they span to higher $B$. All these features suggest that the carbon sensor
signal correlates to the response of the 2DES to MW irradiation.

In Fig. 3 we present the MW-power dependence of $R_{xx}$ and $R_{S}$  versus $B$. The frequency of MW is
$48$ GHz.
The MW-power is swept from 3.2 mW to dark. Remarkably, we obtain similar evolution for both types of measurements:
 for all curves the radiation response is
 getting lower  as the MW-power decreases. It is also
interesting to highlight that the oscillations phase for $R_{xx}$ does not
change with the MW-power. And  it happens the same with $R_{S}$ whose oscillations
phase
remains unchanged for different MW powers.
In other words, as MW power increases the amplitude of the
oscillatory features becomes bigger for both $R_{xx}$ and $R_{S}$,
but their relative position on the x-axis are fixed.
We have used the above results to confirm further the radiation-mediated correlation
between the 2DES and the detected signal in the remote carbon resistor.
Therefore, in Fig. 4 we exhibit $\Delta R_{S}=R_{S}(E_{ST})-R_{S}(0)$ vs
the MW power $P$, where $R_{S}(E_{ST})$ is the calculated $R_{S}$ for the irradiated carbon sensor
and $R_{S}(0)$ without irradiation. As in experiments, we obtain that
the power dependence  of $\Delta R_{S}$ oscillations for
all frequencies can be fit with a sublinear power law function
$\Delta R_{S}=\Theta P^{\alpha}$, where $\Theta$ and $\alpha$ are
fit parameters. The parameter $\Theta$ vary with the MW-frequency but
for the exponent we expect that $\alpha\simeq 0.5$, as it is obtained
in the experiments.

The explanation can be obtained straightforward from the theory.
The radiation power, $P$, can
be related with $A$ through the well-known formula that gives
radiation intensity $I$ (power divided by surface) in terms of the radiation electric field
$E_{0}$: $I=\frac{1}{2}c\epsilon_{0}E_{0}^{2}$, where $c$ is the
speed of light in vacuum and  $\epsilon_{0}$ is the  permittivity in
vacuum. If we want to express only the power in terms of the radiation
electric field we have to take into account the sample surface. In the particular case of GaAs
 we can readily obtain:
$ P = \frac{1}{2} c_{GaAs} \epsilon \epsilon_{0}E_{0}^{2} S$
where $c_{GaAs}$ is the speed of light in GaAs and $\epsilon$
is the GaAs dielectric constant.  Accordingly,
$E_{0}\propto
\sqrt{P}$ and in turn $E_{S0}\propto
\sqrt{P}$ too.
 Then, eventually  we
obtain that $R_{S}$ varies with $P$ following an square root law:
$R_{S}\propto \sqrt{P}$.
Thus, we expect that
the exponent of the sublinear law will be $\alpha\simeq 0.5$.

In Fig. 5, we present the dependence on the dc-current of the MW-response in
the longitudinal voltage $V_{xx}$ and $R_{S}$ versus $B$. The MW-power
remains unchanged. In Figs. 5a and 5b we describe similar magnetoresistance oscillations
as in previous figures for $R_{xx}$  and $R_{S}$. Figs. 5c and 5d correspond to
a situation where the applied dc-current is switched off. As expected, the
$V_{xx}$ MW response vanishes. However, the radiation-induced variation of $R_{S}$
still persists when the MW illumination continues to be applied to the 2DES.
Remarkably, this indicates that there is a MW effect on the 2DES even in
the absence of dc current according to the obtained response in the carbon
resistor. This result demonstrates also the strong correlation between
the irradiated 2DES and the carbon resistor. We have to stand out that
$R_{S}$ turns out to be immune to the applied current through the sample.
In Fig. 6 we present in the same panel, calculated $R_{xx}$ under
MW-irradiation of high power (left $y$-axis)
and $R_{S}$ (right $y$-axis)  versus $B$. Here we want to contrast
the $R_{S}$ response when $R_{xx}$ undergoes a regime of ZRS.
We observe that the radiation-induced features in the $R_{S}$ curve
are totally unaffected by the existence of ZRS in $R_{xx}$.
Radiation-assisted ZRS is mainly a transport-related effect, however
it does not cause any effect on how the radiation is re-emitted by the 2DES.
Then, this  result is in agreement  with the one exhibited in Fig.5, in the
 the sense that the re-emission process from the 2DES is
 totally unaffected by the transport process through the sample.
 To date, this kind of experiment has not been carried out yet.
We think that the corresponding experimental results would
mean a possible way to validate and give credibility to the existing theoretical models
on MIRO and ZRS.

\section{Conclusions}
In conclusion, we have theoretically analyzed recent experiments on re-emission of radiation  in
photoexcited two-dimensional electron systems.
These experiments have shown that there exist a strong correlation
between the re-emitted radiation and the radiation-driven current through the sample.
We study these experimental results using the radiation-driven electron orbits model.
According to it, two-dimensional electrons under a magnetic field move back and forth  driven by
radiation. Therefore, they behave as oscillating dipoles re-emitting part of the radiation
previously absorbed.
 Then, this theory would explain in a simple way the experimental results.
We also studied the dependence of re-emission on MW-frequency, MW-power and dc-current through the 2DES.
In the case of the power dependence we obtain a sublinear power law where
the exponent is around 0.5 as in the experiments. We are able to trace
this exponent and power law using the theoretical model.
We also obtain calculated results on re-emission and transport trough the carbon
resistor when the 2DES undergoes a regime of ZRS. We predict that this regime does not affect
the process of re-emission from the 2DES.

\section{Acknowledgments}

This work is supported by the MCYT (Spain) under grant
MAT2011-24331 and ITN Grant 234970 (EU).

\end{document}